\documentclass{article}

\usepackage{hyperref}
\usepackage[style=numeric-comp,sorting=none]{biblatex}
\usepackage{graphicx}
\usepackage{amsmath}
\usepackage{amssymb}
\usepackage{tensor}
\usepackage{url}
\usepackage{authblk}
\usepackage{tikz}

\newcommand{\sd}{\mathrm{d}}

\title{Practical predictions for the effects of acceleration on decay}
\author[1]{Wim Beenakker}
\author[1]{David Venhoek}
\affil[1]{Institute for Mathematics, Astrophysics and Particle Physics, Radboud University
Nijmegen, Heyendaalseweg 135, Nijmegen, The Netherlands}
\date{December 1, 2025}

\allowdisplaybreaks

\begin{document}

\maketitle

\begin{abstract}
We make predictions of the effects of acceleration on decay rate for three benchmark processes spanning alpha, beta and isomeric decay. These processes are observed to require lower acceleration than that required by previously studied processes. In particular for the case of isomeric decay of Thorium 229m the effects of acceleration are found to be tantalizingly close to being observable with next generation particle accelerators.
\end{abstract}

The Unruh effect, first described in the 70s by Unruh \cite{Unruh} and Fulling \cite{Fulling}, predicts that accelerated observers will perceive a background of particles even when moving through a Minkowski Vacuum. Due to its close relations to Hawking radiation, it has drawn considerable attention, both from a theoretical perspective as well as with an eye towards potential observation. Crispino et. al. provide a relatively recent review \cite{Crispino}.

Early investigations into the possibility of observing the Unruh effect go back as far as the early 80s, by Bell and Leinaas \cite{bell1,bell2}. However, direct observation runs into trouble because of the significant accelerations needed ($10^{30}\frac{\text{m}}{\text{s}^2}$), and the potential for the observational signature to be obscured by thermal effects. However, claims of observation have been made based on data from the NA63 experiment by Lynch et. al.\cite{Lynch1, Lynch2, Lynch3}.

These limitations have caused attention to shift to more indirect ways of observing the Unruh effect. The effect of background thermal baths instead of a vacuum have been studied in \cite{Garay}, and entanglement of two accelerated detectors in \cite{Zhou}.

The change of the decay rate in the rest frame of unstable particles is also one such avenue, first studied by M\"uller \cite{Muller}. Predictions were made for the decay of muons, pions and protons, indicating observable effects could be expected for accelerations around $10^{29} \frac{\text{m}}{\text{s}^2}$ to $10^{30} \frac{\text{m}}{\text{s}^2}$ for the first two. This was expanded upon by \cite{Matsas}, although that study is somewhat hard to interpert due to the use of a 4-dimensional Fermion model in a 2-dimensional setting.

Throughout this paper we will focus more on nuclear decay processes, an approach first taken in \cite{AcceleratingDecay}. Because these processes involve significantly less energetic decays, one can hope, and indeed find, that the required acceleration for observable effects is also significantly lower.

The tools developed in \cite{DetectorEquivalence} have opened up significant freedom in the processes that can be explored numerically, allowing us to make predictions for realistic decay processes in full 4-dimensional spacetime, without needing to neglect masses of decay products to keep analytical calculations tractable. We do need to neglect the effects of recoil, though. It turns out that for many nuclear decays, especially the lower energy ones we are interested in here, recoil is indeed a small effect.

In this paper, we present predictions for three benchmark processes, with the aim of investigating the potential for observability in the near future. We have chosen to focus on one alpha decay (polonium-210), one beta decay (tritium) and one isomeric decay (thorium-229m). Each of these processes were chosen as they are a relatively low-energy example of their type of decay, providing a good indication of the amount of acceleration needed for each of these types. All calculated decay rates are calculated in the rest frame of the decaying particle, making the effects calculated here an intrinsic change in the decay rate, rather than the result of special relativistic effects.

\section{Decay rates of accelerating nuclei}

Throughout the rest of this paper we will be using the main results from \cite{DetectorEquivalence}. This allows us to calculate the decay rate of accelerated nuclei using their decay rate as a function of the mass difference $\Delta = \left(m_i - m_f\right)c^2$, expressed as an energy, between initial and final state nucleus. Here $m_i$ is the mass of the initial nuclei, and $m_f$ the mass of the final state nucleus. Note that $\Delta$ is subtly different from the excess energy in a decay, as the masses of the other decay products are not yet substracted.

Restricting to linear acceleration of magnitude $a$, formulas 10 and 13 from \cite{DetectorEquivalence} combined give (with $\hbar$ and $c$ restored)
\begin{align}
\Gamma_a\left(\Delta\right) &= \frac{1}{2\pi}\int_0^\infty \sd \Delta' \;\; \Gamma_0\left(\Delta'\right) \frac{4c}{a\hbar}e^{\frac{\pi\Delta c}{a\hbar}}K_{\frac{2i\Delta c}{a\hbar}}\left(\frac{2\Delta' c}{a\hbar}\right)\label{eq:transform}
\end{align}
where $\Gamma_0(\Delta')$ is the decay rate as a function of the energy difference for stationary nuclei, and $\Gamma_a(\Delta)$ the accelerated decay rate as function of energy difference, at acceleration $a$. The function $K_\nu(z)$ is the modified bessel function of the second kind, as defined in formula~10.25.3 of \cite{DLMF}.

Furthermore, the results from section 5 of \cite{DetectorEquivalence}, with some generalisations, imply the stationary decay rate can be calculated through the ordinary machinery of quantum field theory, so long as we assume the initial and final state nuclei to be significantly more massive than the other decay products. This allows us to bring in standard results from the theoretical treatment of nuclear physics where necessary.

Thus, for each of the processes of interest in this study, our procedure for calculating their decay rates as function of acceleration becomes relatively straightforward. First, we find an expression for the decay rate of the nuclei as a function of the mass difference between initial and final state nuclei. Then, we plug this into Equation~\ref{eq:transform} given above to get the accelerated decay rate. This is then evaluated, either numerically or analytically, to provide the decay rate as a function of acceleration. Finally, we provide comparison to the acceleration that could, in theory, be provided by the LHC and FCC for the nuclei in question, based on information from \cite{fcchh}.

\section{Numerical evaluation of the accelerated decay rate}

For two of the processes we will discuss, the accelerated decay rate is evaluated numerically. This is implemented as a striaghtforward integration via Simpson's rule, using the archt library \cite{archt} to evaluate the Bessel functions with complex index. The code used for the integration and plotting of the results is available at \url{https://github.com/davidv1992/fourierdecay}.

Numerical errors were constrained to 1 percent relative error on the change in decay rate. This was evaluated by repeating the calculation with half the precision for the Bessel functions, and half the number of integration points and checking that the result of the calculation was within half a percent of the primary calculation.

\section{Alpha decay: $^{210}\text{Po}$}

We start with the analysis of the decay of polonium-210 into lead-206 and an alpha particle. This was chosen for analysis as it is a pure alpha decay, and is relatively low energy for an alpha decay with its excess decay energy being only $5.407$ MeV\footnote{This excludes the mass of the alpha particle, which is included in $\Delta$.} \cite{AME2020}. The decay is modeled as a scalar point interaction between 3 fields, the polonium-210 nucleus, the lead-206 nucleus and the expelled alpha-particle. This allows us to follow the main example in \cite{DetectorEquivalence}.

\begin{figure}
\center{\includegraphics[width=0.8\textwidth]{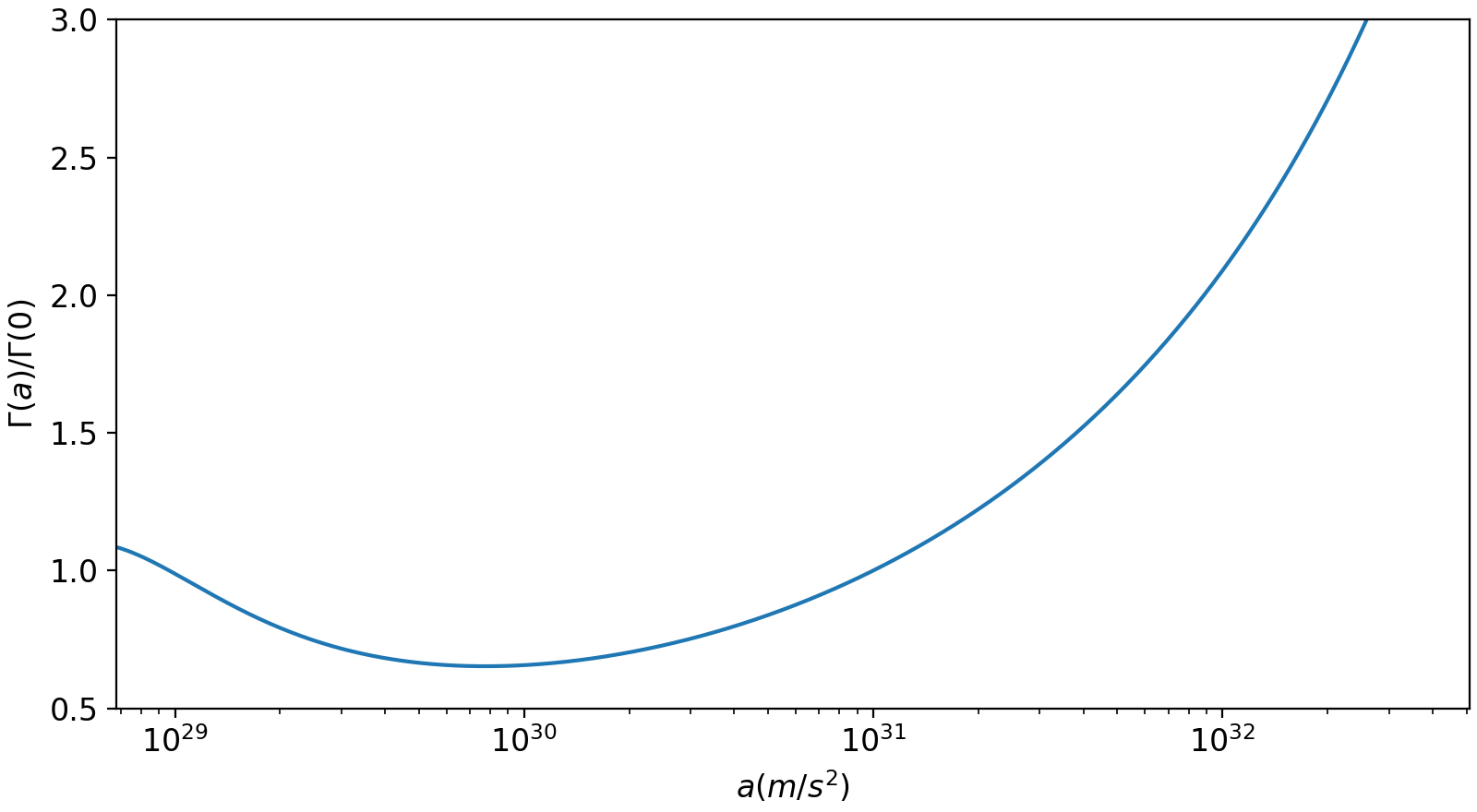}}
\caption{Relative decay rate as function of acceleration for polonium-210}
\label{fig:alphadecay}
\end{figure}

Our choice of model leads to a straightforward constant matrix element for the decay process. Therefore the decay rate as a function of energy difference is primarily determined by the phase space integral, giving:
\begin{align}
\Gamma_0 = \left.\frac{g^2p}{8m_i\pi}\frac{1}{\sqrt{p^2+m_f^2c^2}+\sqrt{p^2+m_\alpha^2c^2}}\right|_{p=c\frac{\sqrt{\lambda(m_i^2,m_f^2,m_\alpha^2)}}{2m_i}}
\end{align}
where $g$ is some constant, $m_i$ is the mass of the polonium-210 nucleus, $m_f$ the mass of the lead-206 nucleus and $m_{\alpha}$ the mass of the alpha particle. The momentum variable $p = \left|\vec{p}\right|$ is the momentum of the alpha particle in the rest frame of the heavy nuclei.

We will assume no recoil, e.g. that the momentum of the alpha particle is small when compared to the mass of the lead-206 nucleus. Doing so corresponds to a limit where $m_i \rightarrow \infty$ and $m_f \rightarrow \infty$ whilst keeping their difference $\Delta = (m_i - m_f)c^2$ constant, simplifying the decay rate to
\begin{align}
\Gamma_0 = \frac{G^2}{2\pi}\sqrt{\Delta^2c^{-4} - m_\alpha^2}
\end{align}
with $G = \frac{g}{2m_f}$.

Using this result, and an alpha particle mass of $3727 \frac{\text{MeV}}{c^2}$\cite{CODATA}, to find the decay rate as a function of acceleration gives Figure~\ref{fig:alphadecay}. Unfortunately, details in the implementation of the calculation of the Bessel functions limit the calculation to accelerations above about $5.9 \cdot 10^{28} \frac{\text{m}}{\text{s}^2}$.

Within the range calculable, we can see the beginnings of an oscillatory pattern on the low-acceleration end, just about exceeding a 10 percent change in decay rate on the peak visible. Note however, these accelerations are significantly above what is likely to be reachable experimentally, as the LHC is only capable of reaching an acceleration of about $2.9\cdot 10^{20} \frac{\text{m}}{\text{s}^2}$ for fully ionized polonium, and the FCC as currently designed reaching $3.9\cdot 10^{21} \frac{\text{m}}{\text{s}^2}$.

\section{Beta decay: $^{3}\text{H}$}

\begin{figure}
\center{\includegraphics[width=0.8\textwidth]{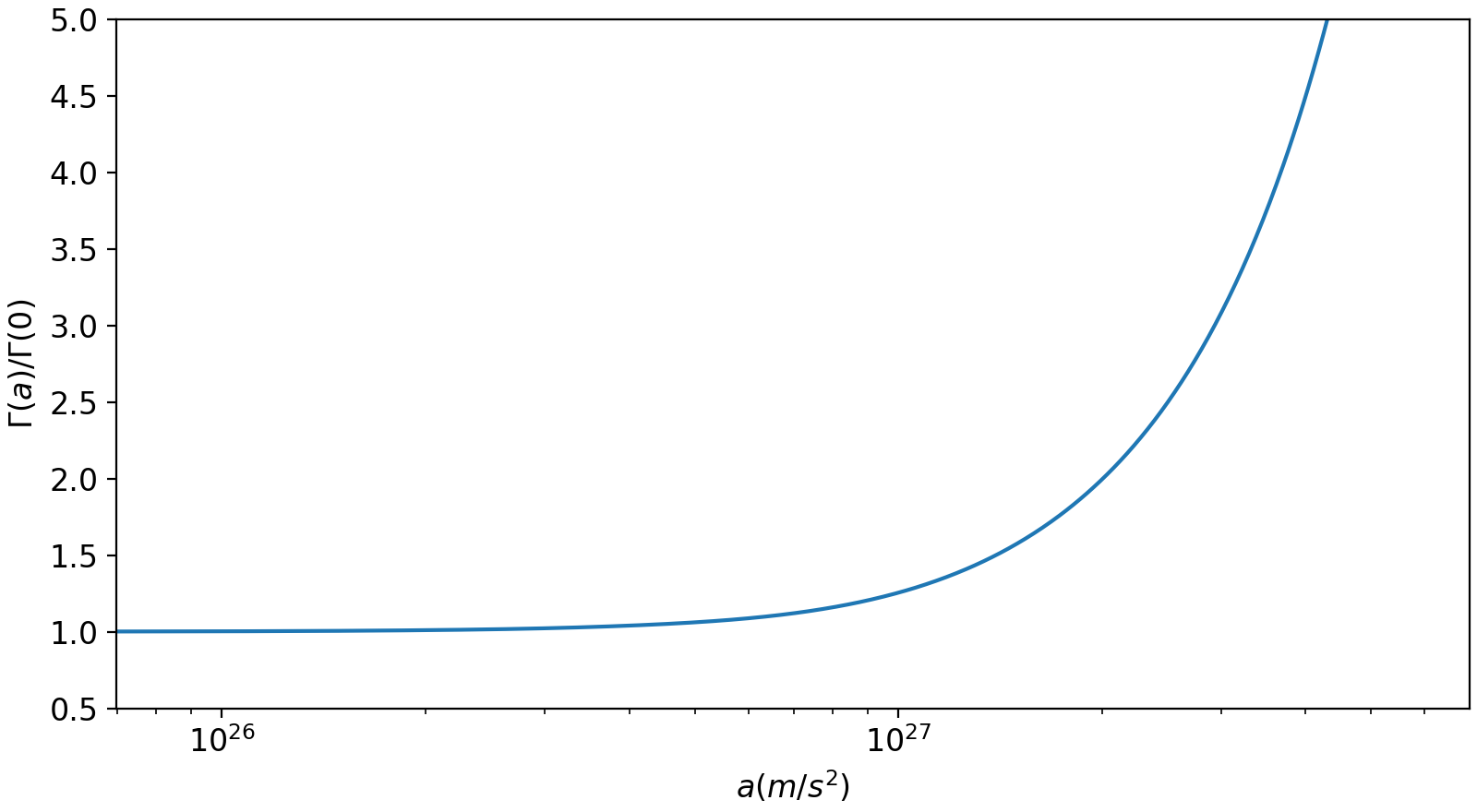}}
\caption{Relative decay rate as function of acceleration for tritium}
\label{fig:weakdecay}
\end{figure}

The decay of tritium into helium-3, an electron, and a neutrino provides a low energy pure beta decay process. For the energy dependence of the decay rate of the tritium, we use the derivation in chapters 45-46 of \cite{Walecka}. At $18.592$ keV excess decay energy \cite{AME2020}\footnote{This excludes the mass of the electron and neutrino, which are included in $\Delta$}, a long wavelength approximation is valid. To leading order, the differential decay rate is parameterized by (equation 46.41 in \cite{Walecka})
\begin{align}
\sd\Gamma_0 &= p_e^2\left(\Delta - c\sqrt{m_e^2c^2+p_e^2}\right)^2\sd p_e \frac{\sd \Omega_e}{4\pi}\frac{\sd\Omega_\nu}{4\pi}\left(C_1\left(1+\frac{\vec{p_\nu}\cdot\vec{p_e}}{\sqrt{p_\nu^2\left(m_e^2c^2+p_e^2\right)}}\right)\right.\nonumber\\
&\phantom{=} \quad+ \left.C_2\left(1-\frac{\vec{p_\nu}\cdot\vec{p_e}}{3\sqrt{p_\nu^2\left(m_e^2c^2+p_e^2\right)}}\right)\right)
\end{align}
where $\vec{p_e}$ is the momentum of the electron, with $p_e = \left|\vec{p_e}\right|$ its norm, $\vec{p_\nu}$ the momentum of the neutrino, with $p_\nu = \left|\vec{p_\nu}\right|$ its norm, $m_e$ the electron mass, and $\Delta$ the mass difference between the helium-3 and tritium nuclei, expressed as an energy. $C_1$ and $C_2$ are constants determined by both the electroweak coupling constants, as well as the detailed structure of the nuclei.

By symmetry, evaluating the phase space integrals in the above simplifies the decay rate significantly to
\begin{align}
\Gamma_0 = C\int_0^{\sqrt{\Delta^2c^{-2}-m_e^2c^2}} p_e^2\left(\Delta - c\sqrt{m_e^2c^2+p_e^2}\right)^2\sd p_e
\end{align}
where $C$ is a combination of the above constants $C_1$ and $C_2$. The simplification to a single proportionality constant is possible as the only difference between the $C_1$ and $C_2$ contributions to the decay is the angular distribution of the decay rate, not its proportionality to the available energy.

The above derivation neglects several effects. Primarily, it is assumed that the wavelength associated with $\Delta$ is large compared to the sizes of the nuclei. This assumption also allows us to assume that the recoil is negligible, and to ignore further terms in the multipole expansion, which is done in the above equations. Furthermore, the mass of the neutrino is ignored, which is reasonable as it is much smaller than $\Delta$, being bound by laboratory-based direct observations to be smaller than $0.45 \text{eV}c^{-2}$ \cite{KatrinNeutrinoMass}.

To calculate the decay rate of tritium, we use the electron mass $510.999 \text{keV}c^{-2}$ \cite{CODATA}, and the excess (excluding electron mass) decay energy for tritium $18.592$ keV \cite{AME2020}. Figure~\ref{fig:weakdecay} shows the decay rate of tritium as a function of acceleration, relative to the decay rate at no acceleration. A 1 percent difference in decay rate occurs at an acceleration of $2.07\cdot 10^{26} \frac{\text{m}}{\text{s}^2}$. For comparison, a rough approximation of the acceleration the LHC can achieve for ionized tritium is roughly $2.0\cdot 10^{20} \frac{\text{m}}{\text{s}^2}$, with the FCC as currently designed being able to reach $2.7\cdot 10^{21}\frac{\text{m}}{\text{s}^2}$.

\section{Isomeric decay: $^{229}\text{Th}$}

\begin{figure}[b!]
\center{\includegraphics[width=0.8\textwidth]{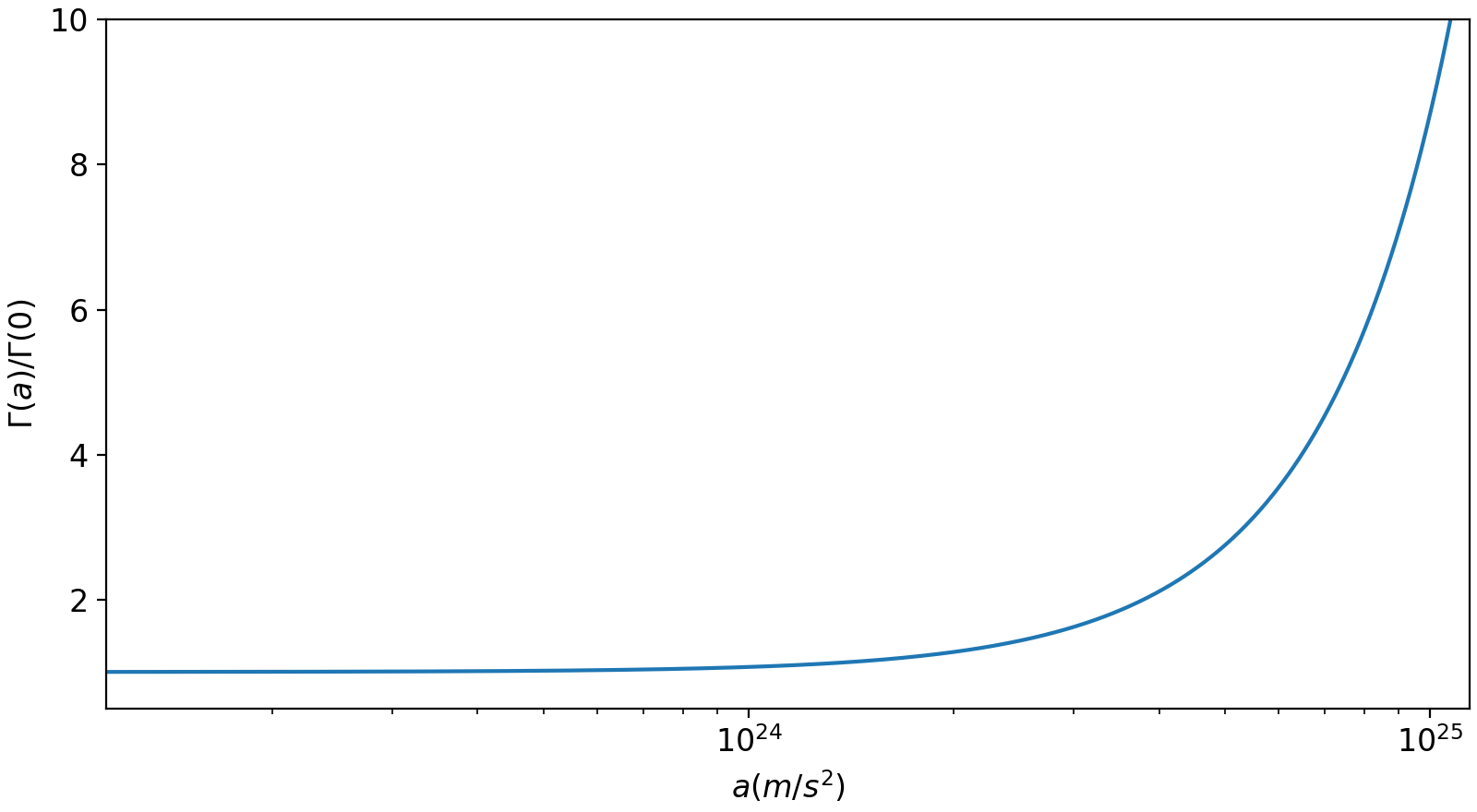}}
\caption{Relative decay rate as function of acceleration for thorium-229m}
\label{fig:photondecay}
\end{figure}

Thorium provides the lowest energy nuclear isomer excitation currently known, recently determined at 8.354eV \cite{ThoriumFrequency}. This very low energy makes it an interesting candidate for detecting the effect of acceleration on decay rate, as it has the potential to change at much lower accelerations than any of the processes considered above.

We will again use the results from \cite{Walecka}, specifically equations 7.44, 7.57 and 7.58 combined to give a decay rate
\begin{align}
\Gamma_0 = C \Delta^3
\end{align}
where $\Delta$ is the energy difference between the initial and final nuclei, and $C$ a proportionality constant determined from the electromagnetic coupling constant and the structure of the nucleus involved.

The above approximation is valid under the assumption that the wavelength associated with $\Delta$ is large with respect to the size of the nuclei. This allows us to ignore higher order contributions in the multipole expansion as well as any recoil of the nucleus emitting the photon.

In this particular case, the decay rate under acceleration can be calculated analytically, using 10.43.19 from \cite{DLMF} together with a number of well-known identities on Gamma functions, yielding
\begin{align}
\Gamma_a = \frac{C \Delta^3}{1-e^{-\frac{2\pi\Delta c}{a\hbar}}}\left(1+\left(\frac{a\hbar}{\Delta c}\right)^2\right)
\end{align}

Using this, the decay rate of accelerated thorium-229m is plotted in Figure~\ref{fig:photondecay}. A 1 percent difference in decay rate occurs at an acceleration of $3.8\cdot 10^{23} \frac{\text{m}}{\text{s}^2}$. For comparison, a rough approximation of the acceleration the LHC can achieve for fully ionized thorium is roughly $2.8\cdot 10^{20} \frac{\text{m}}{\text{s}^2}$, with the FCC as currently designed being able to reach $3.8\cdot 10^{21} \frac{\text{m}}{\text{s}^2}$.

\section{Conclusions}

We have produced predictions for 3 types of decay processes in alpha decay, beta decay and isomeric decay processes. These processes operate at significantly reduced energy scale and as such require significantly less acceleration to see changes in decay rate when compared to earlier studied examples such as muon and proton decay.

Especially the nuclear isomer decay of thorium 299m is showing changes in decay rate at accelerations that are getting tantalizingly close to being reachable with next generation accelerators. Especially should the FCC be realized with stronger magnets, there is significant potential for improvements as the acceleration scales with the square of the beam energy. Unfortunately, the required acceleration as a function of accuracy in measuring the decay rate goes as a square root, with a 4x increase in precision of the measured decay rate only allowing a 2x reduction in the acceleration. Therefore bridging the entire gap is likely to be a significant challenge still.

For future research it would be interesting to study whether effects on these or similar decay rates can be observed in an astrophysical context. Cosmic processes offer the potential for significantly higher accelerations than possible in the lab, though at the cost of more indirect observations and signficantly less control.

Finally, perhaps there are decay or decay-like processes not yet considered here which occur at even lower energy scales, and which are practicable in a labratory setting. The required acceleration for observable effects appears to be directly proportional to the energy released in the decay process, making improvements here an interesting avenue for further lowering the barrier to observation.

\printbibliography

@misc{DLMF,
         key = "{\relax DLMF}",
       title = "{\it NIST Digital Library of Mathematical Functions}",
howpublished = "\url{https://dlmf.nist.gov/}, Release 1.2.2 of 2024-09-15",
         url = "https://dlmf.nist.gov/",
        note = "F.~W.~J. Olver, A.~B. {Olde Daalhuis}, D.~W. Lozier, B.~I. Schneider,
                R.~F. Boisvert, C.~W. Clark, B.~R. Miller, B.~V. Saunders,
                H.~S. Cohl, and M.~A. McClain, eds."}

@article{AME2020,
doi = {10.1088/1674-1137/abddaf},
year = {2021},
month = {mar},
publisher = {Chinese Physical Society and the Institute of High Energy Physics of the Chinese Academy of Sciences and the Institute of Modern Physics of the Chinese Academy of Sciences and IOP Publishing Ltd},
volume = {45},
number = {3},
pages = {030003},
author = {Wang, Meng and Huang, W.J. and Kondev, F.G. and Audi, G. and Naimi, S.},
title = {The AME 2020 atomic mass evaluation (II). Tables, graphs and references*},
journal = {Chinese Physics C}
}

@article{CODATA,
  title = {CODATA recommended values of the fundamental physical constants: 2022},
  author = {Mohr, Peter J. and Newell, David B. and Taylor, Barry N. and Tiesinga, Eite},
  journal = {Rev. Mod. Phys.},
  volume = {97},
  issue = {2},
  pages = {025002},
  numpages = {62},
  year = {2025},
  month = {Apr},
  publisher = {American Physical Society},
  doi = {10.1103/RevModPhys.97.025002},
}

@book{Walecka,
author = {Walecka, John Dirk},
title = {Theoretical Nuclear and Subnuclear Physics},
publisher = {World Scientific and Imperial College Press},
year = {2004},
doi = {10.1142/5500},
address = {},
edition   = {2nd},
}

@article{
KatrinNeutrinoMass,
author = {KATRIN Collaboration† and Max Aker  and Dominic Batzler  and Armen Beglarian  and Jan Behrens  and Justus Beisenkötter  and Matteo Biassoni  and Benedikt Bieringer  and Yanina Biondi  and Fabian Block  and Steffen Bobien  and Matthias Böttcher  and Beate Bornschein  and Lutz Bornschein  and Tom S. Caldwell  and Marco Carminati  and Auttakit Chatrabhuti  and Suren Chilingaryan  and Byron A. Daniel  and Karol Debowski  and Martin Descher  and Deseada Díaz Barrero  and Peter J. Doe  and Otokar Dragoun  and Guido Drexlin  and Frank Edzards  and Klaus Eitel  and Enrico Ellinger  and Ralph Engel  and Sanshiro Enomoto  and Arne Felden  and Caroline Fengler  and Carlo Fiorini  and Joseph A. Formaggio  and Christian Forstner  and Florian M. Fränkle  and Kevin Gauda  and Andrew S. Gavin  and Woosik Gil  and Ferenc Glück  and Steffen Grohmann  and Robin Grössle  and Rainer Gumbsheimer  and Nathanael Gutknecht  and Volker Hannen  and Leonard Hasselmann  and Norman Haußmann  and Klaus Helbing  and Hanna Henke  and Svenja Heyns  and Stephanie Hickford  and Roman Hiller  and David Hillesheimer  and Dominic Hinz  and Thomas Höhn  and Anton Huber  and Alexander Jansen  and Christian Karl  and Jonas Kellerer  and Khanchai Khosonthongkee  and Matthias Kleifges  and Manuel Klein  and Joshua Kohpeiß  and Christoph Köhler  and Leonard Köllenberger  and Andreas Kopmann  and Neven Kovač  and Alojz Kovalík  and Holger Krause  and Luisa La Cascio  and Thierry Lasserre  and Joscha Lauer  and Thanh-Long Le  and Ondřej Lebeda  and Bjoern Lehnert  and Gen Li  and Alexey Lokhov  and Moritz Machatschek  and Martin Mark  and Alexander Marsteller  and Eric L. Martin  and Christin Melzer  and Susanne Mertens  and Shailaja Mohanty  and Jalal Mostafa  and Klaus Müller  and Andrea Nava  and Holger Neumann  and Simon Niemes  and Anthony Onillon  and Diana S. Parno  and Maura Pavan  and Udomsilp Pinsook  and Alan W. P. Poon  and Jose Manuel Lopez Poyato  and Stefano Pozzi  and Florian Priester  and Jan Ráliš  and Shivani Ramachandran  and R. G. Hamish Robertson  and Caroline Rodenbeck  and Marco Röllig  and Carsten Röttele  and Milos Ryšavý  and Rudolf Sack  and Alejandro Saenz  and Richard Salomon  and Peter Schäfer  and Magnus Schlösser  and Klaus Schlösser  and Lisa Schlüter  and Sonja Schneidewind  and Ulrich Schnurr  and Michael Schrank  and Jannis Schürmann  and Ann-Kathrin Schütz  and Alessandro Schwemmer  and Adrian Schwenck  and Michal Šefčík  and Daniel Siegmann  and Frank Simon  and Felix Spanier  and Daniela Spreng  and Warintorn Sreethawong  and Markus Steidl  and Jaroslav Štorek  and Xaver Stribl  and Michael Sturm  and Narumon Suwonjandee  and Nicholas Tan Jerome  and Helmut H. Telle  and Larisa A. Thorne  and Thomas Thümmler  and Simon Tirolf  and Nikita Titov  and Igor Tkachev  and Korbinian Urban  and Kathrin Valerius  and Drahoslav Vénos  and Christian Weinheimer  and Stefan Welte  and Jürgen Wendel  and Christoph Wiesinger  and John F. Wilkerson  and Joachim Wolf  and Sascha Wüstling  and Johanna Wydra  and Weiran Xu  and Sergey Zadorozhny  and Genrich Zeller },
title = {Direct neutrino-mass measurement based on 259 days of KATRIN data},
journal = {Science},
volume = {388},
number = {6743},
pages = {180-185},
year = {2025},
doi = {10.1126/science.adq9592}}

@Article{ThoriumFrequency,
author={Zhang, Chuankun
and Ooi, Tian
and Higgins, Jacob S.
and Doyle, Jack F.
and von der Wense, Lars
and Beeks, Kjeld
and Leitner, Adrian
and Kazakov, Georgy A.
and Li, Peng
and Thirolf, Peter G.
and Schumm, Thorsten
and Ye, Jun},
title={Frequency ratio of the 229mTh nuclear isomeric transition and the 87Sr atomic clock},
journal={Nature},
year={2024},
month={Sep},
day={01},
volume={633},
number={8028},
pages={63-70},
issn={1476-4687},
doi={10.1038/s41586-024-07839-6},
}

@article{fcchh,
  title={{FCC-hh: the Hadron collider: future circular collider conceptual design report volume 3}},
  author={FCC collaboration and others},
  journal={European Physical Journal: Special Topics},
  volume={228},
  number={4},
  pages={755--1107},
  year={2019},
  publisher={Springer Verlag}
}

@misc{DetectorEquivalence,
      title={Unruh detectors, Feynman diagrams, acceleration and decay}, 
      author={Wim Beenakker and David Venhoek},
      year={2025},
      eprint={2501.11516},
      archivePrefix={arXiv},
      primaryClass={gr-qc},
}

@misc{AcceleratingDecay,
      title={Accelerating decay with acceleration}, 
      author={Wim Beenakker and David Venhoek},
      year={2023},
      eprint={2310.06592},
      archivePrefix={arXiv},
      primaryClass={gr-qc},
}

@article{archt, 
    title={Bounds and algorithms for the $K$-Bessel function of imaginary order}, 
    volume={16}, 
    %DOI={10.1112/S1461157013000028}, 
    journal={LMS Journal of Computation and Mathematics}, 
    author={Booker, Andrew R. and Strömbergsson, Andreas and Then, Holger}, 
    year={2013}, 
    pages={78-108}}

@article{Unruh,
  title = {{Notes on black-hole evaporation}},
  author = {Unruh, W. G.},
  journal = {Phys. Rev. D},
  volume = {14},
  issue = {4},
  pages = {870--892},
  numpages = {0},
  year = {1976},
  month = {Aug},
  publisher = {American Physical Society},
  doi = {10.1103/PhysRevD.14.870},
}

@article{Fulling,
  title = {{Nonuniqueness of Canonical Field Quantization in Riemannian Space-Time}},
  author = {Fulling, Stephen A.},
  journal = {Phys. Rev. D},
  volume = {7},
  issue = {10},
  pages = {2850--2862},
  numpages = {0},
  year = {1973},
  month = {May},
  publisher = {American Physical Society},
  doi = {10.1103/PhysRevD.7.2850},
}

@article{Crispino,
  title = {{The Unruh effect and its applications}},
  author = {Crispino, Lu\'{\i}s C. B. and Higuchi, Atsushi and Matsas, George E. A.},
  journal = {Rev. Mod. Phys.},
  volume = {80},
  issue = {3},
  pages = {787--838},
  numpages = {0},
  year = {2008},
  month = {Jul},
  publisher = {American Physical Society},
  doi = {10.1103/RevModPhys.80.787},
}

@article{bell1,
	title = {{Electrons as accelerated thermometers}},
	journal = {Nuclear Physics B},
	volume = {212},
	number = {1},
	pages = {131-150},
	year = {1983},
	issn = {0550-3213},
	doi = {10.1016/0550-3213(83)90601-6},
	author = {J.S. Bell and J.M. Leinaas},
}

@article{bell2,
	title = {{The Unruh effect and quantum fluctuations of electrons in storage rings}},
	journal = {Nuclear Physics B},
	volume = {284},
	pages = {488-508},
	year = {1987},
	issn = {0550-3213},
	doi = {10.1016/0550-3213(87)90047-2},
	author = {J.S. Bell and J.M. Leinaas},
}

@article{Lynch1,
  title = {{Experimental observation of acceleration-induced thermality}},
  author = {Lynch, Morgan H. and Cohen, Eliahu and Hadad, Yaron and Kaminer, Ido},
  journal = {Phys. Rev. D},
  volume = {104},
  issue = {2},
  pages = {025015},
  numpages = {11},
  year = {2021},
  month = {Jul},
  publisher = {American Physical Society},
  doi = {10.1103/PhysRevD.104.025015},
}

@article{Lynch2,
  title={{Experimental observation of a Rindler horizon}},
  author={Lynch, Morgan H},
  journal={arXiv preprint arXiv:2303.14642},
  year={2023}
}

@misc{Lynch3,
      title={Hyperbolic recoil and the Unruh effect at CERN-NA63}, 
      author={Morgan H. Lynch},
      year={2025},
      eprint={2505.21292},
      archivePrefix={arXiv},
      primaryClass={hep-ph},
}

@article{Zhou,
  title={{Detecting circular Unruh effect with quantum entanglement}},
  author={Zhou, Yuebing and Hu, Jiawei and Yu, Hongwei},
  journal={arXiv preprint arXiv:2303.05638},
  year={2023}
}

@article{Garay,
  title = {{Thermalization of particle detectors: The Unruh effect and its reverse}},
  author = {Garay, Luis J. and Mart\'{\i}n-Mart\'{\i}nez, Eduardo and de Ram\'on, Jos\'e},
  journal = {Phys. Rev. D},
  volume = {94},
  issue = {10},
  pages = {104048},
  numpages = {11},
  year = {2016},
  month = {Nov},
  publisher = {American Physical Society},
  doi = {10.1103/PhysRevD.94.104048},
}

@article{Muller,
  title = {{Decay of accelerated particles}},
  author = {M\"uller, Rainer},
  journal = {Phys. Rev. D},
  volume = {56},
  issue = {2},
  pages = {953--960},
  numpages = {0},
  year = {1997},
  month = {Jul},
  publisher = {American Physical Society},
  doi = {10.1103/PhysRevD.56.953},
}

@article{Matsas,
  title = {{Decay of protons and neutrons induced by acceleration}},
  author = {Matsas, George E. A. and Vanzella, Daniel A. T.},
  journal = {Phys. Rev. D},
  volume = {59},
  issue = {9},
  pages = {094004},
  numpages = {9},
  year = {1999},
  month = {Mar},
  publisher = {American Physical Society},
  doi = {10.1103/PhysRevD.59.094004},
}

\end{document}